\documentclass{IEEEcsmag}
\pdfoutput = 1
\usepackage[colorlinks,urlcolor=blue,linkcolor=blue,citecolor=blue]{hyperref}
\usepackage{enumerate}
\usepackage{lipsum}
\usepackage{siunitx}
\usepackage{subcaption}
\usepackage{upmath}
\usepackage{xifthen}
\usepackage{xspace}
\usepackage{wasysym}

\newcommand{\EMAIL}[1]{\href{mailto:#1}{#1}}
\newcommand{\TODO}[1]{{\color{magenta} \textbf{TODO}\ifthenelse{\equal{#1}{}}{\xspace}{:~}#1 }}

\newcommand{\GG}{\mbox{\scshape\small gender-guesser}\xspace}

\newcommand{\COMMITS}{\ensuremath{\mathcal{C}}\xspace}

\jvol{XX}
\jnum{YY}
\paper{Z}
\jmonth{March/April}
\jname{IEEE Software}
\pubyear{2021}

\begin{document}

\title{Gender Differences in Public Code Contributions: a 50-year Perspective}
\author{Stefano Zacchiroli}
\affil{Université de Paris and Inria, France}

\begin{abstract}
  Gender imbalance in information technology in general, and Free/Open Source
  Software specifically, is a well-known problem in the field. Still, little is
  known yet about the large-scale extent and long-term trends that underpin the
  phenomenon. We contribute to fill this gap by conducting a longitudinal study
  of the population of contributors to publicly available software source code.
  We analyze 1.6 billion commits corresponding to the development history of
  120 million projects, contributed by 33 million distinct authors over a
  period of 50 years. We classify author names by gender and study their
  evolution over time.

  We show that, while the amount of commits by female authors remains low
  overall, there is evidence of a stable long-term increase in their proportion
  over all contributions, providing hope of a more gender-balanced future for
  collaborative software development.
\end{abstract}

 \maketitle

\label{sec:related}

\chapterinitial{Gender imbalance} in science is an established and well-known
phenomenon: women are underrepresented in STEM~\cite{hill2010whysofew} and even
more so in computing~\cite{margolis2002womencs}. In the field of software
development, Free/Open Source Software (FOSS) projects have been studied from
the perspective of gender imbalance using various approaches.

\emph{Survey-based studies} have repeatedly reported low women participation.
Surveys up to 2003~\cite{david2008fossdevs} reported 95--99\% man dominance in
FOSS; a 2013 survey~\cite{robles2016womeninfoss} observed a ratio of 10\% women
respondents. These surveys targeted FOSS contributors at large (with no
restriction on project affiliation), relied on participant self-selection, and
reached a maximum of several thousand usable responses each. Specific FOSS
communities have also been studied for gender imbalance, e.g.,
Debian~\cite{oneil2016debiansurvey} or KDE~\cite{qiu2010kdewomen}, with similar
results.

\emph{Quantitative studies} of byproducts of collaborative FOSS development
have analyzed selected projects to quantify gender imbalance in: mailing
lists~\cite{kuechler2012genderfoss}, support forums~\cite{vasilescu2014gender},
GitHub teams~\cite{vasilescu2015gender} and pull
requests~\cite{terrell2017gender}. They have confirmed the under-representation
of female contributors in FOSS and also found evidence of measurable biases
against them.

 \label{sec:intro}

\paragraph{Paper contributions}

A piece of knowledge that is still missing is a large-scale analysis of public
code contributions, to establish a global breakdown of contributions by gender
and to verify if long-term trends about gender participation exist in FOSS
development. This paper contributes to fill these gaps. Specifically, we will
address the following research questions:

\begin{enumerate}[\bfseries RQ1.]

\item What is the overall breakdown by gender in contributions and contributors
  to public source code?

\item Is there a long-term trend in the proportion of contributions and
  contributors to public source code by gender?

\end{enumerate}
Answers to these questions will help confirming or disputing past results on
gender imbalance, this time at the unprecedented scale of public code. If
stable trends were to be observed, they might also provide insights about what
to expect in the future, informing policy making.

In order to answer the research questions we conduct a longitudinal study of
the population of contributors to publicly available source code over a period
of 50 years. To that end we retrieve the commits of more than 120 million
collaborative projects from Software Heritage~\cite{swhcacm2018}, totaling 1.6
billion commits. We then classify author names by gender using a
frequency-based approach implemented on top of
\GG~\cite{santamaria2018genderapi}. Finally we aggregate results by authors and
number of commits, and analyze their evolution over time.

\paragraph{Replication package}

A replication package for this paper is available from Zenodo at
\url{https://zenodo.org/record/4140789} (DOI:
\href{http://dx.doi.org/10.5281/zenodo.4140789}{10.5281/zenodo.4140789}).

 \section{DATASET AND METHODOLOGY}
\label{sec:method}

We have retrieved from Software Heritage~\cite{swhcacm2018,
  swh-msr2019-dataset} a snapshot of all the commits the project has archived
until 2020-05-13. It consists of \num{1661391281} commits (1.66\,B), unique by
SHA1 identifier, harvested from about 120 million public projects coming from
major development forges (GitHub, GitLab, etc.) and source code distributions
(Debian, PyPI, NPM, NixOS, etc.).
For each commit we have its identifier, timestamp, and author full name.

\begin{figure}
  \includegraphics[width=\linewidth]{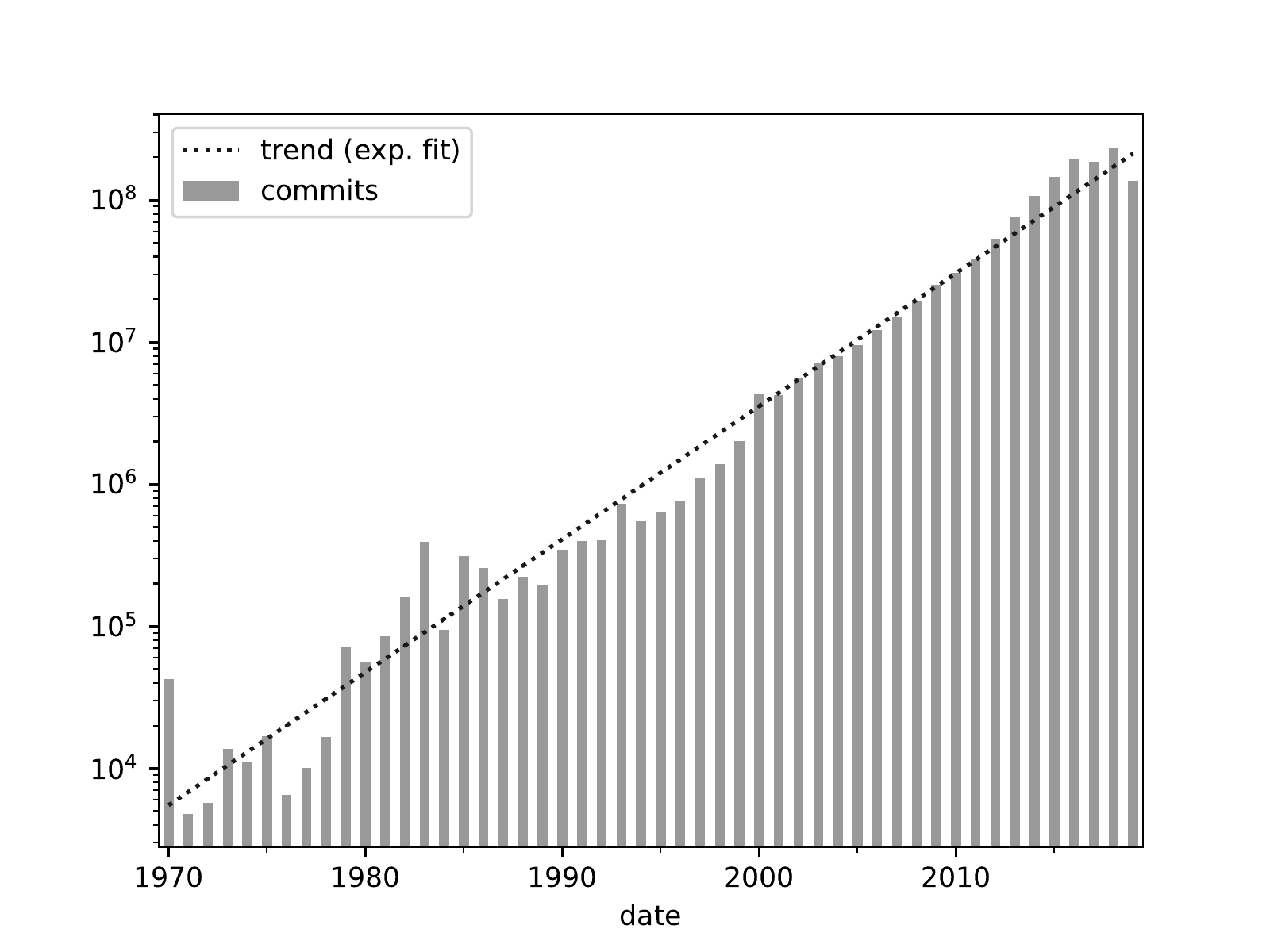}
  \caption{Total number of yearly commits by all authors. The log scale on the
    Y-axis highlights the exponential growth of public code.}
  \label{fig:total}
\end{figure}

We removed from the corpus commits with implausible timestamps, i.e., commits
before the Unix epoch and commits ``in the future'' w.r.t.~the date of the
snapshot, with a tolerance of 1 day. Doing so excluded only 11\,M commits
(0.66\% of the corpus).
Figure~\ref{fig:total} shows the number of commits in the corpus over time. It
exhibits the already observed~\cite{swh-provenance-emse} exponential growth of
public code (the notch for 2020 is a binning artifact due to the incompleteness
of that year in the corpus).

The initial set of \emph{distinct} authors associated to all commits consists
of \num{33660524} (33.7\,M) names. As most version control systems (VCS) do not
store encoding information, author names in the dataset are raw \emph{byte}
sequences. We converted them to Unicode strings, trying the popular UTF-8
encoding and successfully converting \num{33657517} commits (99.991\%).

We then filtered out implausible names such as: email addresses (used by
mistake by authors \emph{in lieu} of their name), names consisting only of
blank characters, overlong names (more than 100 characters), and names
containing more than 10\% non-letter characters. This filtering reduced the
corpus to 26\,M authors after having removed: 7.5\,M non-letter, 150\,K emails,
25\,K blank, and 31 overlong names.
Finally we converted names to lowercase and normalized spaces, obtaining
13.2\,M unique author strings.

Detecting the gender of a name is difficult in
general~\cite{santamaria2018genderapi} and even more so at this scale,
geographic diversity, and lack of curation. Assigning a gender to a name also
reinforces the gender binary, contributing to the marginalization of
individuals who do not identify as men or women. A better approach is to ask
authors for self-identification, but doing so is unfeasible at this scale. We
hence delegate gender inference to automated tooling and we use the results
only in aggregate form to study long-term trends. Throughout the paper we make
no claim about gender identity (as in: the personal sense of one's own gender)
and only discuss gender trends to the extent of which they can be inferred from
author names.

Based on the results of a recent thorough benchmark of gender detection
tools~\cite{santamaria2018genderapi} we have chosen \GG, because it shines on
heterogeneous inputs. \GG is implemented in Python and is open source
(\url{https://pypi.org/project/gender-guesser/}). This last point is
particularly relevant: alternatives based on commercial APIs might give better
accuracy, but would hinder replicability.

\GG takes as input a Unicode string, which is supposed to be a \emph{first}
name, and returns the detected gender as one of 6 possible values, depending on
the tool's certainty about the result: $\{$\textit{male}, \textit{mostly male},
\textit{unknown}, \textit{mostly female}, \textit{female}, \textit{andy}$\}$
(the last one for unisex names).

Authors in our corpus are not split into first v.~family name, but that
distinction is not meaningful anyway in all the world cultures represented in
the corpus~\cite{ishida2011namesaroundtheworld}. Hence, to determine the gender
of an author we apply a \emph{majority criterion}. We use \GG to determine the
gender of each blank-separated \emph{word} in the author name as a string.
Then, if a strict majority of words are detected as belonging to one gender (no
matter how strongly) we associate that gender to the entire author name;
otherwise its gender will remain unknown, formally:
\[
  \begin{array}{l}
    M_a = \{ w\in a | \mathit{guess}(w) \in \{ \textit{male}, \textit{mostly male} \} \} \\
    F_a = \{ w\in a | \mathit{guess}(w) \in \{ \textit{female}, \textit{mostly female} \} \} \\
  \end{array}
\]
where $a$ is an author name from our author corpus, $w$ a word in that string,
and $\mathit{guess}(w)$ denotes the invocation of \GG. The gender of an author
name $a$ is then determined as follows:
\[
  \begin{array}{l}
    \mathit{GG}(a) = \left\{\begin{array}{ll}
                              \mars & \textit{if} \quad |M_a| > |F_a| \\
                              \venus & \textit{if} \quad |F_a| > |M_a| \\
                              ? & \textit{otherwise} \\
                            \end{array}\right.
    \\
  \end{array}
\]

We can now partition the commit corpus \COMMITS in the sets of commits by male
authors, by female authors, or by authors for which we could not determine a
gender, as follows:
\[
  \begin{array}{lcl}
    \COMMITS_{\mars} & = & \{c \in \COMMITS ~|~ \mathit{GG}(c) = \mars \} \\
    \COMMITS_{\venus} & = & \{c \in \COMMITS ~|~ \mathit{GG}(c) = \venus \} \\
    \COMMITS_? & = & \{c \in \COMMITS ~|~ \mathit{GG}(c) = \, ? \} \\
  \end{array}
\]
We have computed these sets in practice, by running \GG on each word in author
names and determining the majority gender for each of them.

 \section{RESULTS}
\label{sec:result}

\begin{figure}
  \centering
  \begin{subfigure}{0.8\linewidth}
    \includegraphics[width=\textwidth]{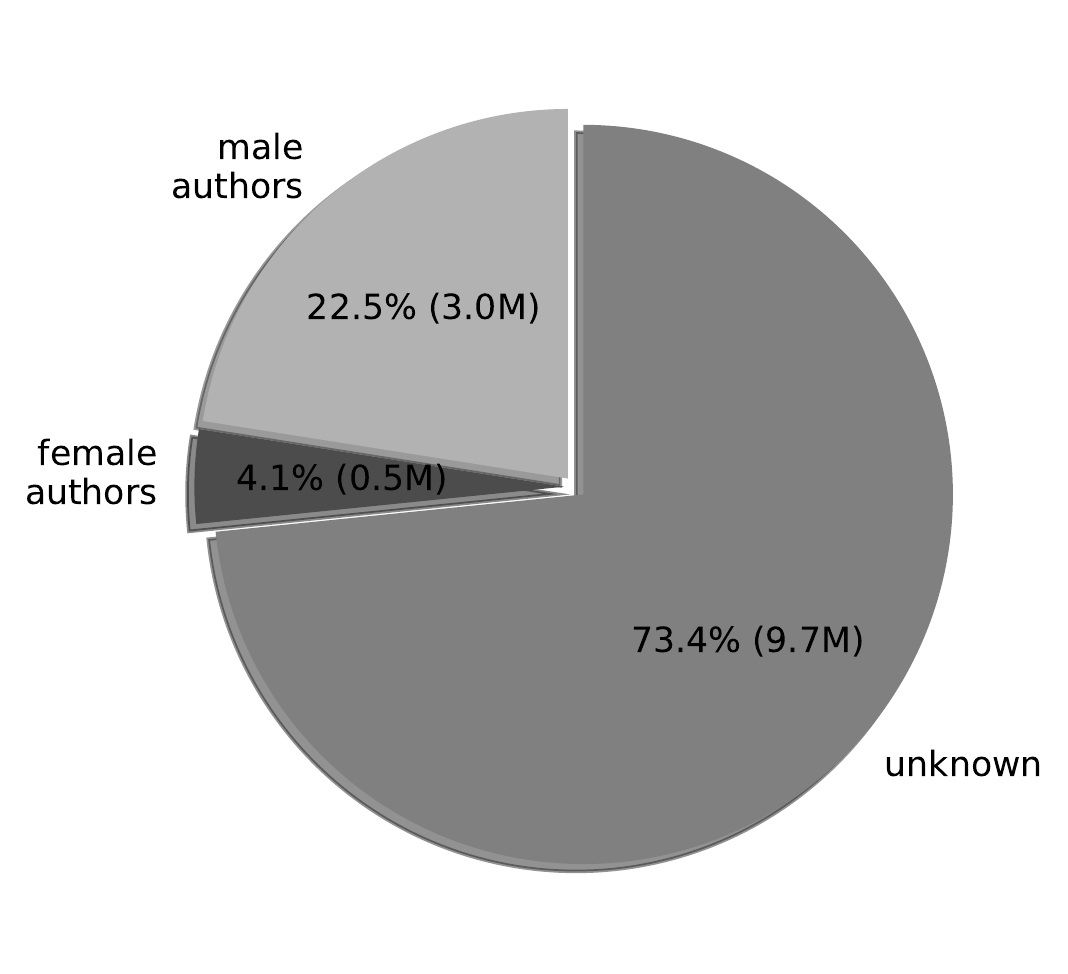}
    \caption{authors}
  \end{subfigure}
  \begin{subfigure}{0.8\linewidth}
    \includegraphics[width=\textwidth]{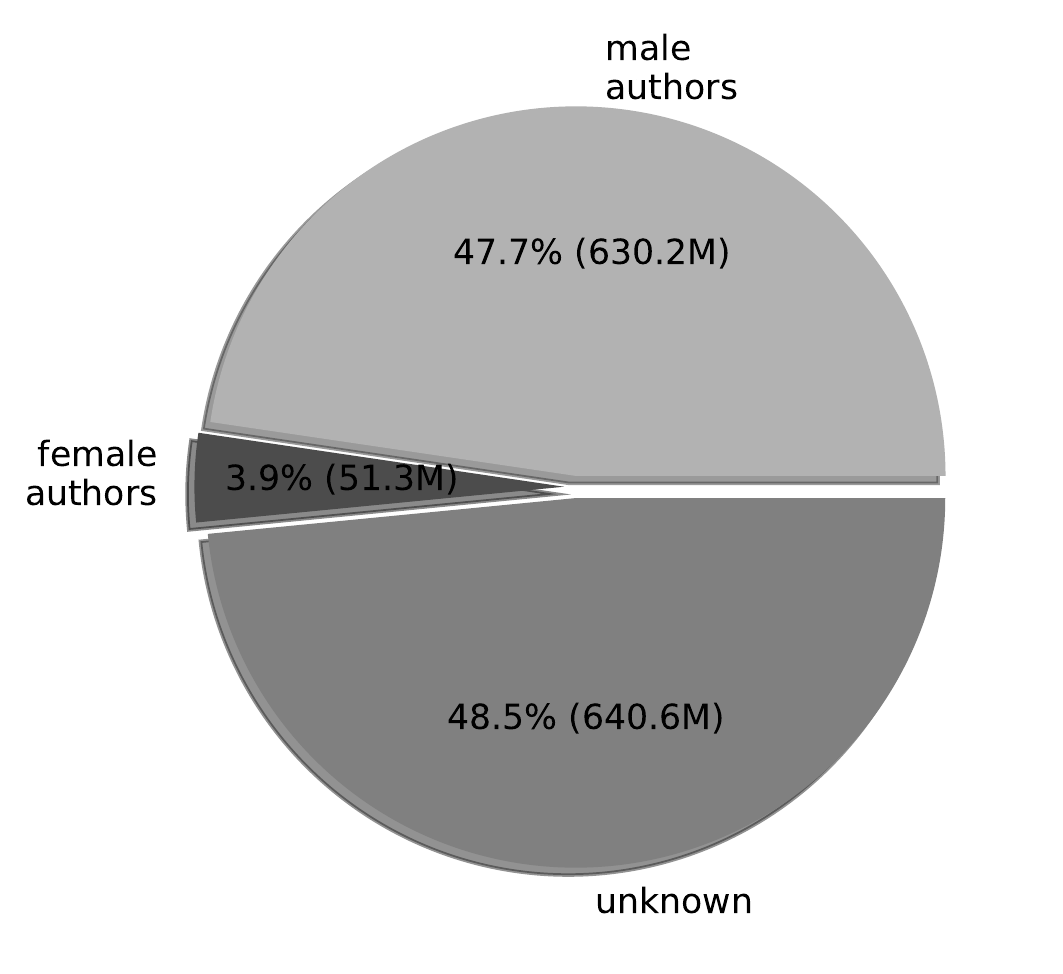}
    \caption{commits}
  \end{subfigure}
  \caption{Breakdown of authors and authored commits by gender}
  \label{fig:total-pie}
\end{figure}

Figure~\ref{fig:total-pie} shows the overall breakdown of detected genders in
the studied corpus (RQ1). We were able to detect a gender for 3.5\,M author
names, or 26.6\% of the author corpus. Author names with a detected gender
account for 682\,M commits, or 51.6\% of the commit corpus. We have verified
that the ratio of commits for which a gender could not be determined remains
within 30--50\% over time. Also, it has been shrinking for the past 20 years,
during which the vast majority of commits have been produced (due to the
exponential growth of the dataset).

\begin{figure*}
  \centering
  \begin{subfigure}{0.495\linewidth}
    \includegraphics[width=\linewidth]{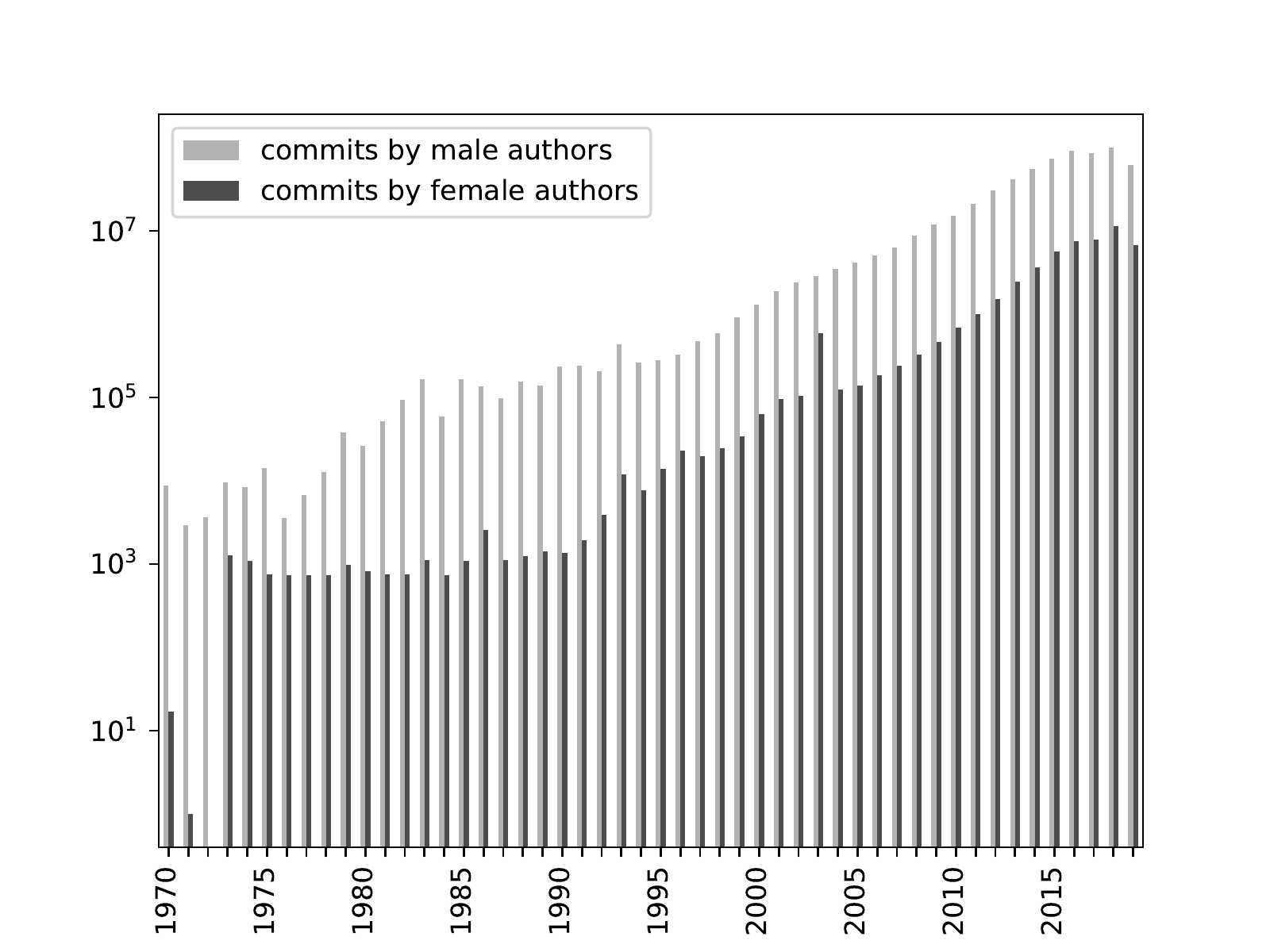}
    \caption{total number of yearly commits by author gender (note the log
      scale on the Y-axis)}
    \label{fig:commits-gender}
  \end{subfigure}
  \begin{subfigure}{0.495\linewidth}
    \includegraphics[width=\linewidth]{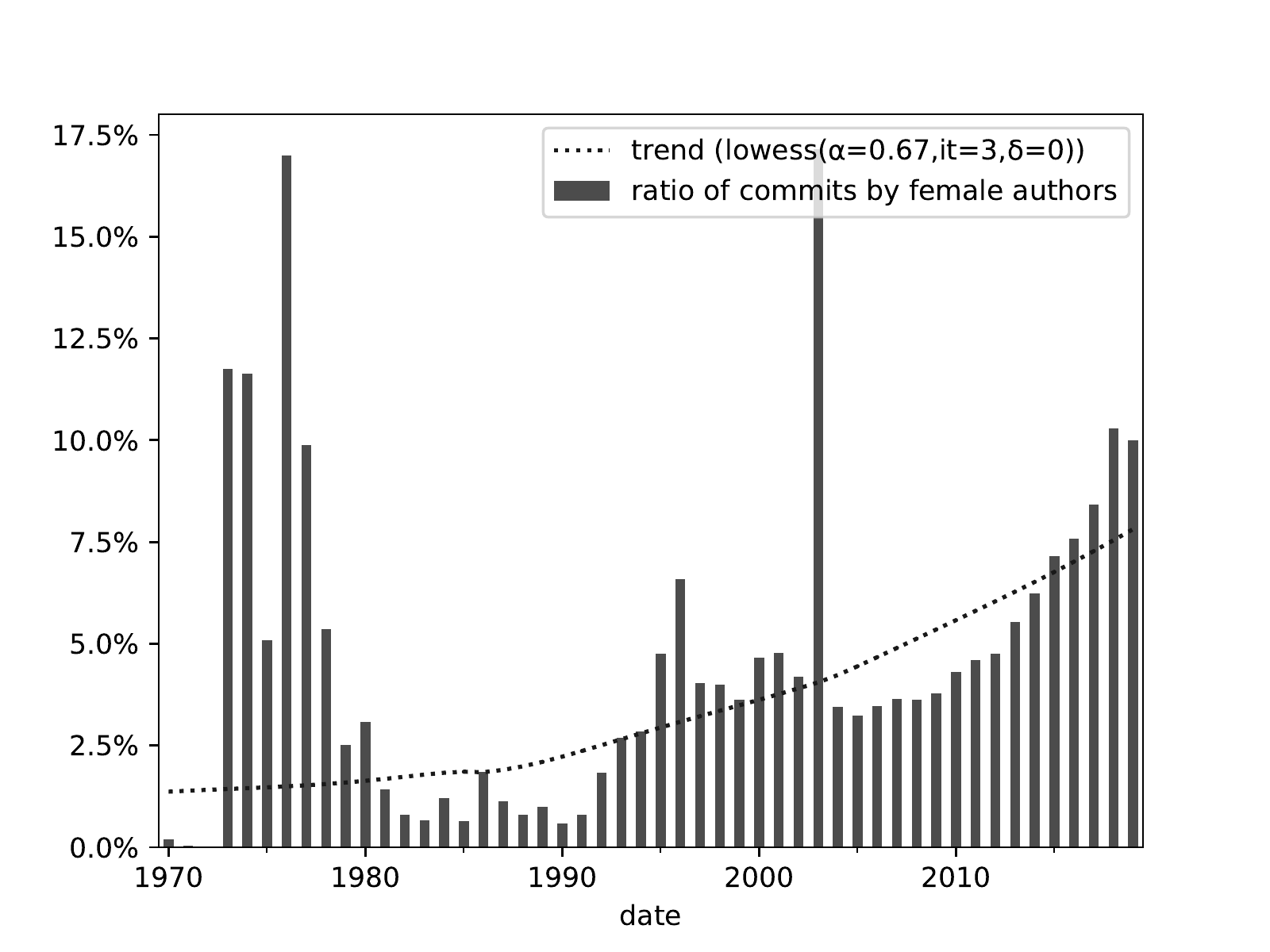}
    \caption{proportion of commits by female authors (on the total of commits
      for which gender could be determined)}
    \label{fig:ratio-female-commits}
  \end{subfigure}
  \caption{Commits breakdown by detected gender over time}
  \label{fig:commits-over-time}
\end{figure*}

Focusing on the author names for which we could determine a gender, 3\,M
(84.6\%) are male authors v.~0.5\,M (15.4\%) female authors. In terms of
contributions, commits by male authors are 630\,M (92.5\% of commits for which
we could determine a gender) v.~51.3\,M (7.5\%) by female authors. In terms of
diversity the picture is pretty dire: \textbf{male authors have contributed
  more than 92\% of public code commits over the past 50 years}.

To answer RQ2, Figure~\ref{fig:commits-gender} shows the evolution over time of
commits authored by gender, excluding commits by authors for which we could not
determine a gender.  Consistently with the breakdown by gender in the corpus as
a whole, we observe that the \emph{yearly totals} of commits by female authors
have lagged behind commits by male authors by significant margins for half a
century.

However, female authors are increasingly contributing to public code.
Figure~\ref{fig:ratio-female-commits} highlights this, showing the 50-year
evolution of the ratio of commits by female authors over the total of commits
for which we could determine a gender. The figure shows both yearly ratios as
percentages and a locally weighted scatterplot smoothing moving regression over
the entire period. \textbf{The ratio of commits by female authors has grown
  steadily over the past 50 years, reaching in 2019 for the first time 10\% of
  all contributions to public code.}

Note also how the growth trend in the ratio of female-authored commits is
steeper over the last 15 years (2005--2019) than before. This is significant
because, due to the exponential growth of public code, those years have
contributed the vast majority of commits to the entire corpus---and hence also
contributed the most to the ongoing ``catch up'' in the total amount of commits
by female authors v.~commits by male authors.

\begin{figure*}
  \centering
  \begin{subfigure}{0.495\linewidth}
    \includegraphics[width=\linewidth]{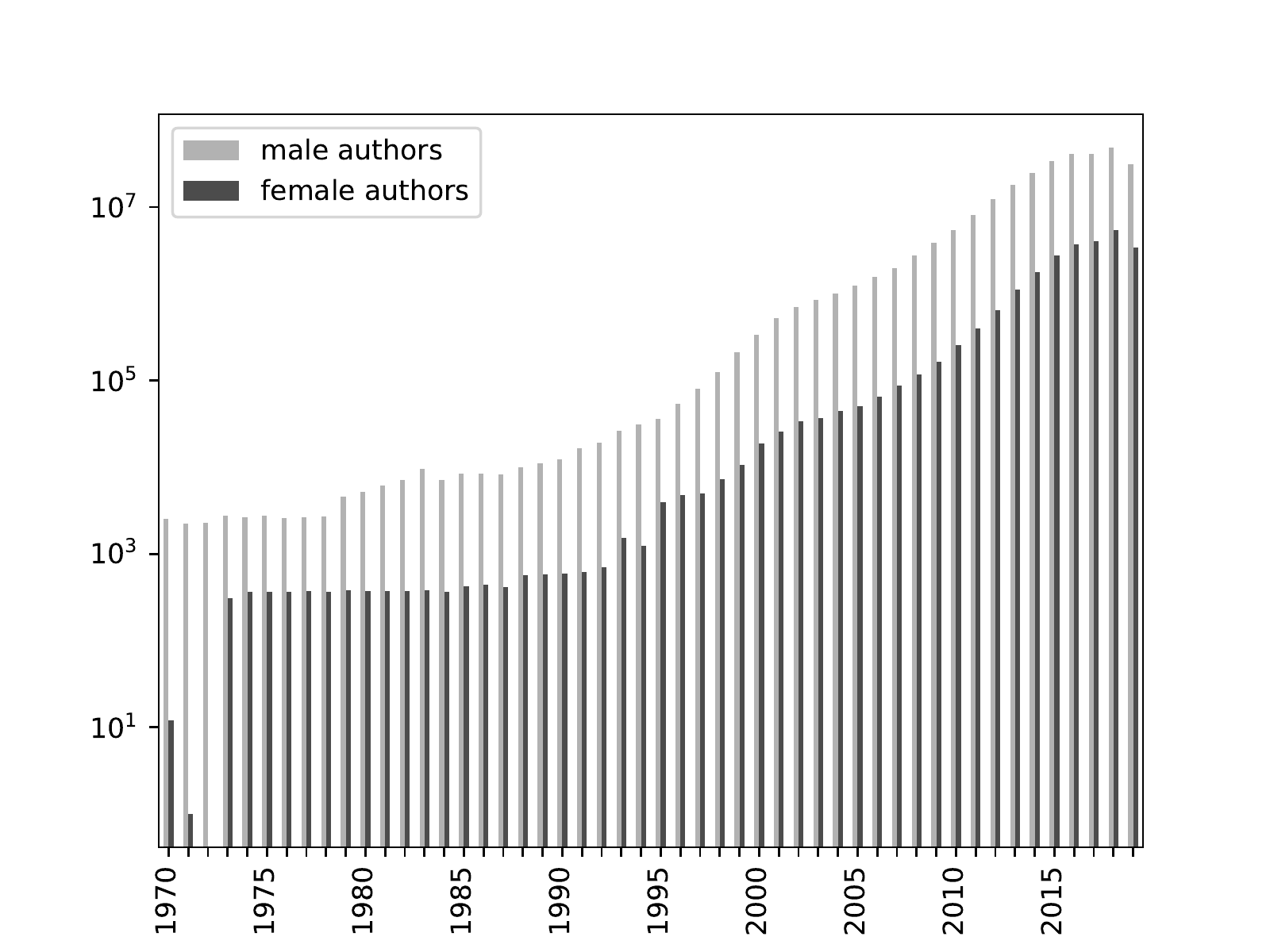}
    \caption{total number of yearly authors by gender (note the log scale on
      the Y-axis)}
    \label{fig:authors-gender}
  \end{subfigure}
  \begin{subfigure}{0.495\linewidth}
    \includegraphics[width=\linewidth]{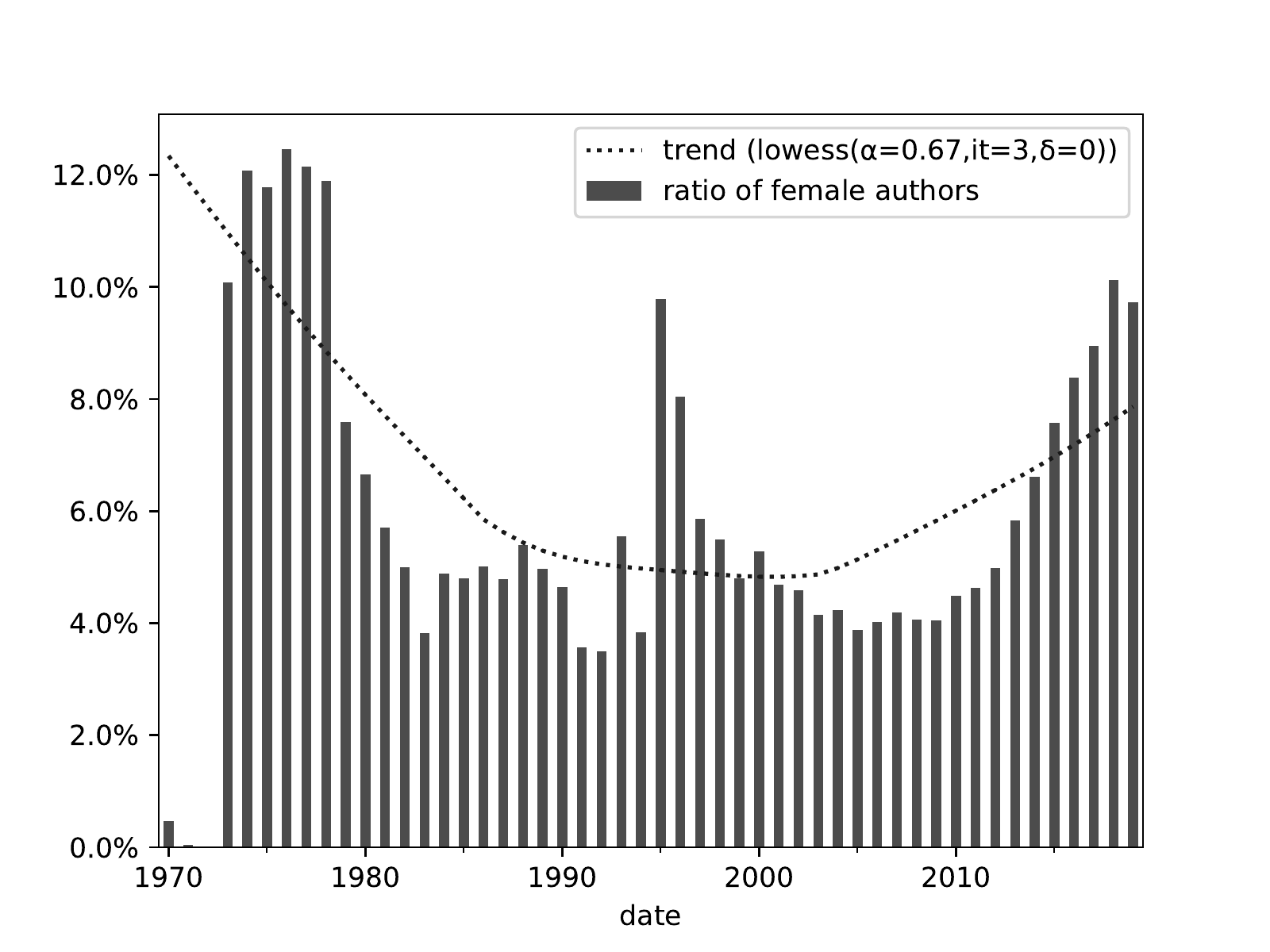}
    \caption{proportion of female authors (on the total of authors for which
      gender could be determined)}
    \label{fig:ratio-female-authors}
  \end{subfigure}
  \caption{Breakdown of authors with at least 1 yearly commit by detected
    gender over time}
  \label{fig:authors-over-time}
\end{figure*}

Figure~\ref{fig:authors-over-time} shows the yearly evolution of the number of
\emph{active authors} by gender, i.e., authors that have contributed at last
one commit in a given year. In particular,
Figure~\ref{fig:ratio-female-authors} confirms the \textbf{significant growth
  of active female authors from around 4\% in 2005 to more than 10\% of all
  public code authors in 2019}. If this trend is to continue, gender diversity
among public code commits authors will increase significantly over the next few
years.

 \section{DISCUSSION}
\label{sec:discussion}
\label{sec:threats}

To the best of our knowledge this is the first longitudinal gender study
performed on public code at this scale---both in terms of population size and
observed time period. The main trade-off in working at this scale is that we
could not rely on curated gender information, e.g., author-provided information
or interviews with them. Also, we had to work with non-parsed author names,
leading to the need of using automated tools and heuristics.

\paragraph{Construct validity}

The approach used for gender detection is crude. It is easy to come up with
examples of family names composed by multiple words that are also common first
names associated to a given gender, which will win majority over the gender
detected for the \emph{actual} first name. We do not expect this to happen
often though. In general, family names are reported as unknown by \GG, not
contributing to shifting the detected author gender in either direction. We
have verified this empirically during experiment design. Doing it more
extensively would boil down to validating the accuracy of \GG itself, which has
already been done in the literature~\cite{santamaria2018genderapi}, supporting
our tool choice.

A related threat is posed by usernames used instead of full names, which happen
in old VCSs like CVS and Subversion. Previous considerations about family names
apply to this case too. Also, 98.2\% of the repositories in Software Heritage
are from VCSs that support author full names (e.g., Git, Mercurial), so we do
not expect this problem to be statistically significant.

Author names for which we could not determine a gender might impact our
results. Given how unknown responses by \GG do not affect gender assignment to
authors, we consider this loss of coverage an acceptable consequence of our
tool choice. As it is customary, we have excluded unknown gender authors from
analysis. We could still determine author gender for almost 700\,M commits,
which remains an unprecedented scale for gender imbalance studies.

Qualitatively, the consistency of the observed trends with recent survey-based
work~\cite{robles2016womeninfoss} about the increase of women participation in
FOSS further supports our results and vice-versa.

\paragraph{External validity}

We do not claim to have analyzed the entire body of collaboratively developed
software. We have nonetheless analyzed the largest publicly available corpus of
commits coming from public version control system repositories. We do not think
a much larger coverage is achievable without, for instance, adding large
non-public forges (e.g., coming from large-scale inner source practices) to our
sample, which would hinder replicability and impact on corpus diversity.

 \section{CONCLUSION}
\label{sec:outro}

We have conducted the first large-scale longitudinal study of gender imbalance
among authors of collaboratively developed, publicly available code.  The study
spans 1.6 billion commits harvested from 120 million projects and contributed
by 33 million authors over a period of 50 years.

Results give a mixed message about gender diversity in public code
collaboration. Overall, contributions by female authors remain scarce: less
that 8\% of commits for which we could detect a gender, confirming decades of
gender imbalance in Free/Open Source Software (FOSS).

On the other hand, contributions by female authors appear to be on the rise and
are rising faster than those by male authors. In 2019 and for the first time in
half a century commits by female authors have reached 10\% of yearly
contributions to public code. Looking at active FOSS authors over time we find
evidence of a similar sustained growth and increasing speed. If the trend of
the past 15 years is to continue, FOSS authors and their contributions might
soon reach a level of gender diversity comparable to other fields.

\medskip

The goal of this study was, on purpose, broad and longitudinal. As future work
we intend to maintain the longitudinal angle, but drill down into specific
software ecosystems to check if significant differences in gender participation
trends exist and, if so, why. Our results also hint at other differences in
participation by gender---e.g., commits per person over time and weekly
participation patterns---which we also intend to explore in future work.

 \section{ACKNOWLEDGMENTS}

The author would like to thank Jesus M. Gonzalez-Barahona for insightful
discussions on this study, as well as Molly de Blanc and Antoine Pietri for
comments on early versions of this paper.

 \begin{IEEEbiography}{Stefano Zacchiroli}{\,}is Associate Professor of Computer Science at Université de Paris on leave at
  Inria, France. His research interests span formal methods, software
  preservation, and Free/Open Source Software engineering. He is co-founder and
  CTO of the Software Heritage project.
Contact him at \EMAIL{zack@irif.fr}.
\end{IEEEbiography}

\end{document}